\documentstyle[osa,manuscript]{revtex}

\begin{document}
\title{
%\hspace*{4.3in} {\normalsize YITP-00-7}\\ 
%\vspace*{-0.2in}
%\hspace*{4.3in} {\normalsize January 2000}\\
Mass modification of $D$-meson at finite density
in QCD sum rule}
\author{Arata Hayashigaki
\footnote{Present address: \\
Department of Physics, Faculty of Science,
University of Tokyo, Tokyo 113-0033, Japan.\\
E-mail address: arata@nt.phys.s.u-tokyo.ac.jp 
or arata@yukawa.kyoto-u.ac.jp}
}
\address{Yukawa Institute for Theoretical Physics, 
Kyoto University, Kyoto 606-8502, Japan}
\maketitle
\begin{abstract}
We evaluate the mass shift of isospin-averaged $D$-meson
in the nuclear medium.
Borel-transformed QCD sum rules are used to describe 
an interaction between the $D$-meson and a nucleon 
by taking into account all the lowest dimension-4 operators 
in the operator product expansion (OPE).
We find at normal matter density
the $D$-meson mass shift is about $10$ times ($\sim 50$ MeV) larger
than that of $J/\psi$.
This originates from the fact that the dominant contribution
in the OPE for the $D$-meson is the nucleon matrix element of 
$m_c\bar{q}q$, where $m_c$ is the charm-quark mass and $q$
denotes light quarks.
We also discuss that the mass shift of the $D$-meson in nuclear matter 
may cause the level crossings of the charmonium states
and the $D\bar{D}$ threshold.
This suggests an additional mechanism of the $J/\psi$ suppression
in high energy heavy-ion collisions.

\vspace*{0.5in}

\noindent
PACS numbers: 12.38.Lq, 13.85.-t, 14.40.Lb, 21.65.+f

\noindent
Keywords: QCD sum rule, $D$-meson, mass modification,
nuclear medium, $D$-$N$ interaction, $J/\psi$ suppression

\end{abstract}
\section{Introduction}
Changes of hadron properties in the nuclear medium 
have recently attracted great interests in theoretical studies.
They also have induced 
the on-going experiments and forthcoming experimental plans
in heavy nuclei in GSI and in relativistic 
heavy-ion collisions in SPS at CERN and AGS, RHIC at BNL\cite{QM97}.
In particular the spectral changes of vector mesons are expected to be 
a possible signal to investigate such medium effect,
because their leptonic decay in the medium 
can supply us information of the medium 
without disturbance of the strong interaction.
Motivated by this experimental advantage,
the in-medium effect of the light vector-mesons ($\rho^0,\omega$ and $\phi$) 
has been mainly studied in effective hadronic models\cite{BR} 
and QCD sum rules (QSR's)\cite{HL,YK,JL,KH,FKW,FHL,HKL}. 
Recently we applied the QSR method to low-lying heavy-quarkonium ($J/\psi$)
in order to investigate
its mass shift in nuclear matter and 
the $J/\psi$-nucleon interaction at low energy\cite{AH1}.
It was shown 
that the $J/\psi$ mass at normal matter density
drops by $0.1\sim 0.2 \%$ of its vacuum value. 
This negative mass-shift of
$J/\psi$ corresponds to a small decrease of the mass, 
$4\sim 7$ MeV\cite{AH1,KKLMW}. 
This is much smaller than the similar effect in the light
mesons such as $\rho^0$ and $\omega$ \cite{HL,YK,JL,KH}.
This difference can be understood as follows: 
In QSR the magnitude of the spectral change is related to the
in-medium change of quark and gluon condensates \cite{CFG,JCFG}. 
In the linear density approximation, the quark operators give
a dominant contribution for $\rho^0$ and $\omega$, 
and have significant change in the matter.
On the other hand,
the dominant in-medium gluon condensates 
at the normal matter density
for $J/\psi$
are only $5\sim 10\%$ smaller than its vacuum value. 

In this paper, we generalize the above calculations and
apply the QSR analysis in Ref.\onlinecite{KH} 
to light and heavy quark systems with unequal mass.
We focus on in-medium properties of pseudoscalar mesons $D$'s.
Studying the mass modification of the $D$-mesons in the 
nuclear medium is important
by the following physical reasons:
\begin{enumerate}
\item
A part of total cross section of the $D$-meson production
attributes a conversion of the $c\bar{c}$-states such as $J/\psi$ 
and $\psi'$ into open-charm pairs. The component is produced  
through the reaction, $\psi+N \rightarrow D+\bar{D}$,
in high energy proton-nucleus collisions and 
relativistic high energy heavy-ion collisions.
It contains a heavy charm-quark and a light quark. 
The existence of a light quark in the $D$-meson causes
much difference of the properties in the nuclear medium
between the $D$-meson and $J/\psi$.
The latter is dominated by the gluon condensates as discussed above, 
while the former has a large contribution 
from the light-quark condensates, multiplied by a charm-quark mass.
Furthermore $J/\psi$ predominantly interacts with light hadrons solely
through gluonic content effects in matter, while the $D$-mesons
will couple strongly through inelastic channels such as 
$DN$ $\rightarrow$ $\Lambda_c$ or $\Sigma_c$ $(+\pi)$.
Thus we can expect a large modification of the $D$-meson
in the nuclear medium as well as $\rho^0$ and $\omega$.
\item
The existence of novel states, so-called 
the $D$-mesic nuclei, has been predicted by quark meson coupling (QMC) model
\cite{TLTSL}. The QMC model suggests
that $D^-(\bar{c}d)$ may be bound in heavy atoms 
such as $^{208}$Pb
by an attractive scalar-meson exchange and 
an attractive Coulomb force.
It also suggests that $D^0(c\bar{u})$ is deeply
bound in the nuclei by an attractive $\omega$-meson exchange.
In view of QSR, it is of importance to investigate 
the $D$ meson-nucleon ($N$) interaction.
\item 
The nuclear absorption of the $D$-mesons created 
in $\pi$-$A$ and $p$-$A$ collisions
($A$ denotes targets such as Be, Cu, Al, W and Au)
has been measured via the decay $D\rightarrow K\pi$ 
at Fermilab\cite{Alves}. The result indicates that $D$-mesons are
not completely absorbed in the nuclear targets
irrespective of the charge of the produced $D$-mesons.
The theoretical analysis for the $D$-$N$ interaction
will give an important suggestion for the experimental data
in this case too.
\end{enumerate}
Motivated by these points, we 
investigate the properties of $D$-mesons in the nuclear matter 
through the $D$-$N$ interactions, using an application
of Borel-transformed QSR to the $D$-$N$ forward scattering amplitude.
Here we deal with isospin-averaged $D$-meson current
for simplicity.

After a brief explanation of the 
QSR formulation, in the next section,
we calculate the mass shift at finite density using the Borel QSR
and compare the results with the mass shift of $J/\psi$ calculated 
in Ref.\onlinecite{AH1}. 
On the basis of these results,
we will discuss a possible mechanism
of $J/\psi$ suppression 
through the level crossings between the charmonium states 
and the $D\bar{D}$ threshold.

\section{QCD sum rule analysis for $D$-meson mass shift}
We start with a two-point in-medium correlation function 
$\Pi_{PS}^{\rm NM}$
to discuss hadron properties in the nuclear matter.
In the Fermi gas approximation for the matter,
$\Pi_{PS}^{\rm NM}$ is divided into two parts
by applying the operator product expansion (OPE) to the correlators 
in the deep Euclidean region ($Q^2=-q^2>0$).
One is a vacuum part, $\Pi_{PS}^{\rm 0}$, and another is 
a static one-nucleon part, $T_{PS}$.
This decomposition is expected to be valid at relatively
low density\cite{DL}.
Namely, in the framework of QSR,
$\Pi_{PS}^{\rm NM}$ can be approximated 
reasonably well in
the linear density of the 
nuclear matter that all nucleons are at rest
\cite{KH,CFG,DL}:
\begin{eqnarray}
\Pi_{PS}^{\rm NM}(q) = 
i\int d^{4}x\ e^{iq \cdot x}
\langle \mbox{T}J_5(x)J_5^{\dag}(0) \rangle_{{\rm NM}(\rho_N)}
\simeq \Pi_{PS}^{0}(q)+
\frac{\rho_N}{2M_N}T_{PS}(q),
\label{eqn:2.1}
\end{eqnarray}
where $\rho_N$ denotes the nuclear matter density, and
the forward scattering amplitude $T_{PS}$
of the pseudoscalar current-nucleon is 
\begin{eqnarray}
T_{PS}(\omega,\mbox{\boldmath $q$}\,) =
i\int d^{4}x e^{iq\cdot x}\langle N(p)|
\mbox{T}J_5(x)J_5^{\dag}(0) |N(p) \rangle.
\label{eqn:2.2} 
\end{eqnarray}
Here $q^{\mu}=(\omega,\mbox{\boldmath $q$}\,)$ is the four-momentum carried by
the $D$-meson current, $J_5(x)$ $=$ $J_5^\dag(x)$ $=$
$(\overline{c}i\gamma_5 q(x)+\overline{q}\gamma_5 c(x))/2$,
where $q$ denotes $u$ or $d$ quark.
$|N(p)\rangle$ represents the isospin, spin-averaged
static nucleon-state with the four-momentum
$p = (M_N,\mbox{\boldmath $p$}={\bf 0})$, where $M_N$ is the nucleon mass,
$0.94$ GeV. The state is normalized covariantly as 
$\langle N(\mbox{\boldmath $p$})|N(\mbox{\boldmath $p$}')\rangle = (2\pi)^{3}
2p^{0}\delta^{3}(\mbox{\boldmath $p$}-\mbox{\boldmath $p$}')$. 
The QSR analysis on the forward scattering amplitude
enables us to obtain the information for the $D$-$N$
interaction. In Eq.~(\ref{eqn:2.1}), the second term means a slight
deviation from in-vacuum properties of the $D$-meson determined by
$\Pi^{\rm 0}_{PS}$.
By applying QSR to $T_{PS}$, we get
the $D$-$N$ scattering length $a_D$ in the limit of 
$\mbox{\boldmath $q$}\rightarrow {\bf 0}$.
In this limit, $T_{PS}$ can be related to
the $T$-matrix, 
${\cal T_{DN}}(m_D,\mbox{\boldmath $q$}={\bf 0})$ $=$
$8\pi(M_N+m_D)a_D$. Near the pole position of the $D$-meson,
the spectral function $\rho(\omega,\mbox{\boldmath $q$}={\bf 0})$ is given 
with three unknown phenomenological parameters $a,b$ and $c$ in terms of
the $T$-matrix:
\begin{eqnarray}
\rho(\omega,\mbox{\boldmath $q$}
={\bf 0}) &=& -\frac{1}{\pi}\frac{f_D^2m_D^4}{m_c^2}
\mbox{Im} \left[\frac{{\cal T_{DN}}(\omega,{\bf 0})}{(\omega^{2}-
m_D^2+i\varepsilon)^{2}} \right]
+ \cdots
\label{eqn:2.3} \\
&=& a\,\delta'(\omega^{2}-m_D^2) + b\,\delta(\omega^{2}-m_D^2)
 + c\,\delta(\omega^{2}-s_{0}).
\label{eqn:2.4}
\end{eqnarray}
Here the leptonic decay constant $f_D$ is defined
by the relation $\langle 0|J_5|D(k)\rangle = f_Dm_D^2/m_c$,
where $|D(k)\rangle$ is the $D$-meson state 
with the four-momentum $k$. 
$m_D=1.87$ GeV and $m_c=1.35$
GeV are masses of the $D$-meson and charm quark respectively.
The terms denoted by $\cdots$ in Eq.(\ref{eqn:2.3}) represent 
the continuum contribution and $\delta'$ in Eq.(\ref{eqn:2.4}) 
is the first derivative of $\delta$ function
with respect to $\omega^2$. The first term proportional to $a$
is the double-pole term 
corresponding to the on-shell effect of the $T$-matrix and $a$
is related to the scattering length $a_D$ as
$a=-8\pi(M_N+m_D)a_D f_D^2m_D^4/m_c^2$.
The second term proportional to $b$ 
is the single-pole term corresponding to the off-shell
effect of the $T$-matrix. The third term proportional to $c$
is the continuum term 
corresponding to other remaining effects, where $s_{0}$ is the 
continuum threshold in the vacuum. 
Combining the single-pole
term of $\Pi_{PS}^0$ in Eq.~(\ref{eqn:2.1}) with Eq.~(\ref{eqn:2.4}),
we can relate the scattering length extracted from the QSR of $T_{PS}$
with the mass shift of the $D$-meson,
\begin{eqnarray}
\delta m_D =
2\pi\frac{M_{N}+m_D}{M_Nm_D}\rho_N a_D.
\label{eqn:2.5} 
\end{eqnarray}
We may determine these unknown parameters by 
using a dispersion relation and matching the phenomenological 
(ph) side with the OPE side. 
Before such analysis, we can
impose a constraint from the low energy theorem among these parameters:
In the low energy limit ($\omega\rightarrow 0$), 
$T_{PS}(\omega,{\bf 0})$ becomes
equivalent to the Born term $T_{PS}^{\rm Born}(\omega,{\bf 0})$,
$T_{PS}^{\rm ph}(0)=T_{PS}^{\rm Born}(0)$.
We assume two cases to take the Born term into the ``ph'' side.
The case (i) and the case (ii) are defined as follows\cite{KH}.
At $q_\mu \neq 0$, we require the $\omega^2$-dependence of the Born term
explicitly in the case (i):
\begin{eqnarray}
T_{PS}^{\rm ph}(\omega^2)=T_{PS}^{\rm Born}(\omega^2)
+\frac{a}{(m_D^2-\omega^2)^2}+\frac{b}{m_D^2-\omega^2}+
\frac{c}{s_0-\omega^2},
\label{eqn:2.5p} 
\end{eqnarray}
with the condition
\begin{eqnarray}
\frac{a}{m_D^4}+\frac{b}{m_D^2}+
\frac{c}{s_0}=0.
\label{eqn:2.5pp}
\end{eqnarray}
In the case (ii), we do not require the $\omega^2$-dependence 
of the Born term explicitly:
\begin{eqnarray}
T_{PS}^{\rm ph}(\omega^2)=
\frac{a}{(m_D^2-\omega^2)^2}+\frac{b}{m_D^2-\omega^2}+
\frac{c}{s_0-\omega^2},
\label{eqn:2.6p} 
\end{eqnarray}
with the condition
\begin{eqnarray}
\frac{a}{m_D^4}+\frac{b}{m_D^2}+
\frac{c}{s_0}=T_{PS}^{\rm Born}(0).
\label{eqn:2.6pp}
\end{eqnarray}
If $T_{PS}^{\rm Born}(0)=0$, two cases coincide.
We determine two unknown 
phenomenological parameters 	
$a$ and $b$ in the QSR after the parameter $c$ is removed by the condition
of Eq.~(\ref{eqn:2.5pp}) or (\ref{eqn:2.6pp}).
The analysis of the Born term is easily performed
through a calculation of the Born diagrams at the tree level.
The isospin states of the $D$-meson determine the contribution to  
the Born term as follows:
For the $D^0(c\bar{u})$-$N$ and $D^+(c\bar{d})$-$N$ interactions,
we need two reactions,
\begin{eqnarray}
D^0(c\bar{u})+p(uud)\ \mbox{or}\ n(udd)
&\longrightarrow& \Lambda_c^+,\Sigma_c^+(cud)\ \mbox{or}\ 
\Sigma_c^0(cdd)
\label{eqn:2.8p}
\end{eqnarray}
and 
\begin{eqnarray}
D^+(c\bar{d})+p(uud)\ \mbox{or}\ n(udd)
&\longrightarrow& \Sigma_c^{++}(cuu)\ \mbox{or}\ 
\Lambda_c^+,\Sigma_c^+(cud),
\label{eqn:2.8}
\end{eqnarray}
where the static masses of charmed baryons are $2.28$ GeV for $\Lambda_c^+$
and $2.45$ GeV for $\Sigma_c^0$, $\Sigma_c^+$ and $\Sigma_c^{++}$.
Note that we use $M_B\sim 2.4$ GeV as the average value,
where $B$ means either $\Lambda_c^+$, $\Sigma_c^+$, $\Sigma_c^{++}$
or $\Sigma_c^0$.
After averaging over the nucleon spin, we obtain
\begin{eqnarray}
T_{PS}^{\rm Born}(\omega,{\bf 0})=\frac{1}{2}\frac{M_N(M_N+M_B)}
{\omega^2-(M_N+M_B)^2}
\ g_p(\omega)^2.
\label{eqn:2.7}
\end{eqnarray}
The pseudoscalar form-factor $g_p(q^2)$ is introduced through
the following relation \cite{NN},
\begin{eqnarray}
\langle B(p')|\bar{Q}i\gamma_5 q|N(p)\rangle 
= g_p(q^2)\bar{u}(p')i\gamma_5 u(p),
\label{eqn:2.7p}
\end{eqnarray}
where $u(p)$ is a Dirac spinor and $Q(q)$ in the pseudoscalar current
is the heavy(light)-quark. Furthermore, $g_p(q^2)$
is related to a coupling constant $g_{NDB}$ such as
\begin{eqnarray}
g_p(q^2)=\frac{2m_D^2f_D}{m_q+m_c}\frac{g_{NDB}}{q^2-m_D^2},
\label{eqn:2.7pp}
\end{eqnarray}
where $m_q$ denotes a light-quark mass.
$g_{ND\Lambda_c}$ has been estimated in Ref.\onlinecite{NN}
which gives $g_{DN\Lambda_c}\simeq 6.74$. 
However since $g_{DN\Sigma_c}$ has not been
evaluated, we take an approximation 
$g_{DN\Sigma_c}$ $\simeq$ $g_{ND\Lambda_c}$.
On the other hand, there are no inelastic channels like the above 
for $\bar{D}^0(\bar{c}u)$-$N$ and $D^-(\bar{c}d)$-$N$ interactions, 
i.e. $T_{PS}^{\rm Born}(0)=0$ in both case (i) and case (ii).

Eventually we determine parameters $a$ and $b$
simultaneously after applying the Borel transform to 
both the OPE side and the ``ph'' side.

\section{Numerical results in the Borel sum rule}
After performing the Borel transform $\hat{B}$
of the $T_{PS}$ calculated in the OPE up to dimension-4 \cite{AH2}, 
we obtain as a function of the Borel mass $M$
\begin{eqnarray}
\hat{B}\left[T_{PS}^{\rm OPE}\right]
&=& \frac{1}{2}e^{-m_c^2/M^2}\left[-m_c\langle\bar{q}q\rangle_N
+2\langle q^\dag i D_0 q\rangle_N\left(-1+\frac{m_c^2}{M^2}\right)\right.
\nonumber\\
&+&\frac{1}{2}m_q\langle\bar{q}q\rangle_N
\left(1+\frac{m_c^2}{M^2}\right)+\frac{1}{24}\langle
\frac{\alpha_s}{\pi} G^2\rangle_N
\left(2-\frac{m_c^2}{M^2}\right)
\nonumber\\
&+&\frac{\alpha_s}{\pi} \langle (u\cdot G)^2-\frac{1}{4}G^2 \rangle_N
\frac{1}{3}\left\{
\frac{4}{3}-\frac{1}{6}\frac{m_c^2}{M^2}
+\frac{1}{2}\left(\frac{m_c^2}{M^2}\right)^3\right.	
\nonumber\\
&&+ e^{m_c^2/M^2}
\left(-2\gamma_E-\mbox{\rm ln}\left(\frac{m_c^2}{M^2}\right)
+\int_{0}^{\frac{m_c^2}{M^2}}dt\frac{1-e^{-t}}{t}\right)
\nonumber\\
&&+\left.\left.\left(1-\frac{m_c^2}{M^2}\right)
\mbox{\rm ln}\left(\frac{m_c^2}{4\pi\mu^2}\right)\right\}\right],
\label{eqn:3.1}
\end{eqnarray}
where $\langle \cdot \rangle_N$ denotes the nucleon matrix element.
The renormarization scale is taken to be $\mu^2= 1$ GeV$^2$ and
the Euler constant is $\gamma_E = 0.5772\cdots$.
Corresponding formula for the ``ph'' side, in the case (i) reads
\begin{eqnarray}
\hat{B}\left[T_{PS}^{\mbox{\rm ph}}\right]
&=& a\ \left(\frac{1}{M^2}e^{-m_D^2/M^2}
-\frac{s_0}{m_D^4}e^{-s_0/M^2}\right)
+b\ \left(e^{-m_D^2/M^2}-\frac{s_0}{m_D^2}e^{-s_0/M^2}\right)
\nonumber\\
&+& \frac{1}{2}\frac{M_N(M_N+M_B)}{(M_N+M_B)^2-m_D^2}
\frac{4f_D^2m_D^4}{(m_q+m_c)^2}\,g_{NDB}^2
\nonumber\\
&\times&\left[-\frac{e^{-(M_N+M_B)^2/M^2}}{(M_N+M_B)^2-m_D^2}
+\left(\frac{1}{(M_N+M_B)^2-m_D^2}-\frac{1}{M^2}\right)
e^{-m_D^2/M^2}\right],
\label{eqn:3.2}
\end{eqnarray}
and that in the case (ii) reads
\begin{eqnarray}
\hat{B}\left[T_{PS}^{\mbox{\rm ph}}\right]
&=& a\ \left(\frac{1}{M^2}e^{-m_D^2/M^2}
-\frac{s_0}{m_D^4}e^{-s_0/M^2}\right)
+b\ \left(e^{-m_D^2/M^2}-\frac{s_0}{m_D^2}e^{-s_0/M^2}\right)
\nonumber\\
&&\qquad\qquad + s_0\, T_{PS}^{\rm Born}(0)\,e^{-s_0/M^2}.
\label{eqn:3.2p}
\end{eqnarray}
In Eq.~(\ref{eqn:3.1}), a convention, $(u\cdot G)^2 \equiv 
G_{\kappa\lambda}^a G_{\rho}^{a\;\lambda}\;u^\kappa u^\rho$
is used, where $u=(1,{\bf 0})$ for the static nucleon. 
We equate Eq.~(\ref{eqn:3.1}) with Eq.~(\ref{eqn:3.2}) or 
Eq.~(\ref{eqn:3.2p}) in the case (i) or the case (ii) respectively.
Furthermore we take a first derivative of its equation
with respect to $M^2$ in each case.
We can derive the $D$-meson mass shift as a function of 
$M^2$ by removing the parameter $b$ 
from two equations obtained thus.
The Borel curve for the $D$-meson mass shift, $\delta m_D$, is shown
in Fig.1.
Here we adopt $f_D\simeq 0.18$ GeV for the decay constant
and $s_0\simeq 6.0$ GeV$^2$ for the continuum threshold.
The values of these parameters were
estimated by the analysis of the in-vacuum correlation function in
Ref.\onlinecite{AE,RRY}.
This $f_D$ value is very close to the result of the 
lattice QCD calculation
($0.194\pm 0.01$ GeV)\cite{EKMRS} 
and is consistent with the upper bound of experimental data 
($\leq$ $0.31$ GeV) \cite{PDG}.
Note that the parameters are fixed in our calculation.
We use the nucleon matrix elements \cite{JCFG} 
for quark fields such as
$\langle\bar{q}q\rangle_N\simeq 5.3$ GeV
and $\langle q^\dag i D_0 q\rangle_N\simeq0.34$ GeV$^2$, and 
for gluon fields such as $\langle 
\frac{\alpha_s}{\pi} G^2\rangle_N$ $\simeq$ $-1.2$ 
GeV$^2$ and $\frac{\alpha_s}{\pi}
\langle (u\cdot G)^2-\frac{1}{4}G^2 \rangle_N$
$\simeq$ $-0.1$ GeV$^2$. The light-quark mass is taken to be
$m_q\simeq 0.008$ GeV.
In Fig.1, the solid line (``${\rm Born}=0$''[case (i)])
are calculated without the contribution of inelastic channels.
In the doted line (``With Born term''[case (ii)]) and
the dashed line (``Without Born term''), we allow for 
the contribution of the sub-threshold resonances 
$\Lambda_c$ and $\Sigma_c$, lying very close 
to the $D$-$N$ threshold. 
As shown in Fig.1, the sub-threshold effect is
repulsive both in the case (i) and the case (ii) and 
its contribution in the case (ii) is stronger 
than that in the case (i).
This is analogous to the case for the $K^-$-$p$
interaction \cite{KMN,YNMK}.
In Fig.1, the remaining three lines are evaluated by
extracting the contribution of 
$m_c\langle \bar{q}q\rangle_N$ term in the OPE
from the solid, dashed and doted lines respectively.
We find that the contribution of $m_c\langle \bar{q}q\rangle_N$
term is more than $95\%$ of total contribution
corresponding to the solid, dashed and doted lines
within the plateau regions.

The analysis of this graph is summarized in Table 1.
We cannot determine th Borel windows, since there is 
only lowest-dimension term in the OPE side.
Therefore we take the following procedure to determine a window
of $M^2$:
We focus on a plateau region of each line shown in Fig.1.
First, we take the minimum point in the line as the smallest
value of $\delta m_D$. 
Next, we determine two points in $M^2$
as the deviation from the minimum value of $\delta m_D$
becomes less than $10\%$ of the minimum value. 
We take the region between the two points as a window of $M^2$.
The window of $M^2$ in each line is given in Table 1.
As shown in Table 1, all the Borel curves are rather stable 
within the windows.
Note that the windows discussed above
are also close to that obtained by 
scaling up typical Borel-windows for the light-vector mesons\cite{KH}.
An estimate from the light-vector mesons is
$(\frac{m_D}{m_V})^2\times 0.8\,(=4.7)<M^2
<(\frac{m_D}{m_V})^2\times 1.3\,(=7.7)$ for 
$V=\rho, \omega$ and 
$(\frac{m_D}{m_V})^2\times 1.3\,(=4.4)<M^2
<(\frac{m_D}{m_V})^2\times 1.8\,(=6.1)$ for
$V=\phi$.

It should be stressed that the QSR for vacuum correlation
function $\Pi_{PS}^0$ up to dimension-6 operators
cannot reproduce well 
the $D$-meson mass ($m_D$) in free space\cite{RRY2}.
The Borel curve of $m_D$ at $s_0=6.0$ GeV$^2$
does not have any stability and seems to give
larger values than the experimental data ($m_D=1.87$ GeV)
in the plateau region ($M^2=3 \sim 8$ GeV$^2$) 
of $\delta m_D$ discussed
above. Furthermore, the curve of $m_D$ has 
rather larger change than that of $\delta m_D$ in the region.
Therefore, if we perform the QSR analysis for the effective mass 
$m_D^*=m_D+\delta m_D$, the stability in the Borel curve of $m_D^*$ 
will become obscure. This implies that
the QSR analysis for $m_D^*$ is not so valid in this case.
However the application of the QSR to $\delta m_D$
gives us a possible indication for the $D$-meson mass-shift
as discussed above.

From the above analysis we obtain
$\delta m_D=-48\pm 8$ MeV by considering both cases ((i) and (ii)).
Then the $D$-$N$ scattering length $a_D$ is $- 0.72 \pm 0.12$ fm.
This result suggests that the $D$-$N$ interaction is more attractive
than the $J/\psi$-$N$ interaction, where
$\delta m_{J/\psi}$ is about $-5$ MeV
and $a_{J/\psi}$ is about $- 0.1$ fm \cite{AH1,KKLMW}.
It is noted the above results were performed
at the nuclear matter density, $\rho_N=\rho_0 \sim 2\rho_0$, 
where $\rho_0$ is the normal matter density.
Assuming that
the linear density approximation in the QSR is valid at the 
nuclear matter density,
we find these results lead to a larger decrease of the $D$-meson mass
than that of the charmonium.
We also find that the large mass-shift of the $D$-meson originates
from the contribution of $m_c\langle\bar{q}q\rangle_N$ term
in the OPE.
In the QMC model, the contribution 
from the $m_c\langle\bar{q}q\rangle_N$ term may
correspond to a quark-$\sigma$ meson coupling.
In fact, the model predicts the mass shift of the $D$-meson
becomes $-60$ MeV for the scalar potential at the normal matter density
\cite{TLTSL}. Their results are very close to our results.

\section{Concluding remarks}
We present an analysis of 
isospin-averaged $D$-meson mass-shift through
a direct application of the Borel QSR to the forward scattering amplitude
of the pseudoscalar current and nucleon.
Here we perform the operator product expansion with all the terms
up to dimension-4. The result predicts an attractive mass-shift
of the isospin-averaged $D$-meson about $50$ MeV 
(about $3\%$ of the bare mass) at the normal matter density.
The contribution of $m_c\langle \bar{q}q\rangle_N$ term in the OPE 
is more than $95\%$ of the results evaluated up to
all the dimension-4 operators.
Our result is very close to that reported by the
QMC model\cite{TLTSL}.
Then the $D$-$N$ scattering length $a_D$ is about $-0.7$ fm
and indicates a strongly attractive force between the $D$-meson and
nucleons.
The mass modification of the $D$-meson in the nuclear matter 
corresponds to $10$ times larger than that of $J/\psi$. 

Recent NA50 Collaboration\cite{NA50} 
has reported a strong (so called ``anomalous'') 
$J/\psi$ suppression\cite{MS}
in $Pb$-$Pb$ collision at $158$ GeV per nucleon.
We suggest a following possibility to cause
such a strong $J/\psi$ suppression using the above 
large difference of the mass shifts
between the $D$-meson and $J/\psi$:
Suppose that the other $c\bar{c}$-states such as $\psi'$ and
$\chi_c$'s do not also have a large mass-shift in the nuclear matter
as discussed above for $J/\psi$.
Then on the basis of our calculations,
the $D\bar{D}$ threshold ($\sim 3.74$ GeV in free space) 
decreases by $100$ MeV at $\rho_N=\rho_0$
and comes down between
$\chi_{c2}$ ($\sim 3.55$ GeV) and $\psi'$ ($\sim 3.68$ GeV).
At $\rho_N=2\rho_0$, it decreases by $200$ MeV and comes down
between $\chi_{c1}$ ($\sim 3.50$ GeV) and $\chi_{c2}$ ($\sim 3.55$ GeV). 
Such level crossings between the $D\bar{D}$ threshold
and the charmonium spectrum are shown in Fig.2. 
It is well known from the $p$-$A$ collision data that 
only $60\%$ of $J/\psi$ observed are 
directly produced and the remainder comes from
excited states ($\chi_c$($1P$), $\psi'$($2S$)) with a ratio
of 3 to 1 \cite{Anto}. 
So if the disappearance of $\psi'$ and $\chi_c$ due to
their decays into $D\bar{D}$ takes place, it could
lead to a decrease of the $J/\psi$ yield.
This $J/\psi$ suppression will have stair-shaped form
as a function of the energy density of the system.
At low energy density, only $\psi'\rightarrow$ $D\bar{D}$ 
takes place, which causes a slight suppression of $J/\psi$.
At intermediate density, subsequent suppression occurs
by the level crossing of ($\chi_{c2},\chi_{c1}$) states
with the $D\bar{D}$ threshold.
At higher density, direct suppression of $J/\psi$
by the decay $J/\psi \rightarrow$ $D\bar{D}$ occurs.
This is an alternative or supplementary effect
to the deconfinement scenario in Ref.\onlinecite{KNS}
and may have implications to the recent data 
of NA50 Collaboration\cite{AH3}.
Strictly speaking, as mentioned before,
the formulation of QSR used here
is applicable to dilute
nuclear matter in comparison to highly dense nuclear matter
produced in high energy heavy-ion collision.
If we try to discuss the mass shifts at higher density,
it will need an extension or some improvements
in the above formulation.
This is one of important future works for us.
However we may say that our calculation in this paper
can suggest a possibility of the level crossings
even in only 1 or 2 times normal matter density
and lets one expect the direct $J/\psi$ suppression
at sufficiently higher density.

We may also suggest another phenomenon caused by the level crossings
discussed above.
It is a change of the decay width of the charmonium.
In the vacuum the resonances above $D\bar{D}$ threshold, for example
the $\psi''$ state, have a width of order MeV because of the 
strong open-charm channel.
On the other hand, the resonances below the threshold have a very
sharp width of a few hundreds keV. So after the level crossings,
the decay modes of the $\psi'$ and $\chi_c$ states
will change drastically at least one order of magnitude.

Needless to say, in the high energy heavy-ion collisions
we must also perform the theoretical investigation 
of finite temperature effect \cite{FHL,VJ,STST} 
to the mass modification. 
Therefore, a future task will be to understand
the medium effect in terms of the QSR when 
both temperature and density are finite.
As for investigation of only the $D$-$N$ interaction,
an inverse kinematics experiment
will give useful information\cite{Kar}.
Since the projectile (target) 
is a heavy-ion beam (light-nuclei), the decays of 
the $D$-meson \cite{Alves} and the charmonium inside a nucleus
will be possible in such an experiment.

Before closing, it is stressed that
we should take the above analysis for 
the isospin-averaged $D$-$N$ interaction as a qualitative
estimate. 
If we wish more quantitative discussion,
the isospin-decomposition of the $D$-meson and higher correction terms
beyond dimension-4 operators must be taken into account. 
Particularly the odd-components in the OPE play important roles
for a difference of the mass shift between $D$ and $\bar{D}$.

\section*{Acknowledgements}
I would like to thank T. Hatsuda for careful reading
of the manuscript and many useful discussion.
I also thank S.H. Lee and D. Jido
for helpful discussion and suggestion.

\begin{figure}
%\begin{center}
%\epsfile{file=mshift.eps, scale=1.5}
%\end{center}
\caption{The $D$-meson mass-shift $\delta m_D$ [GeV] 
at normal matter density ($\rho_0=0.17$ fm$^{-3}$)
as a function of the squared Borel mass $M^2$ [GeV$^2$] :
The solid, doted and dashed lines are calculated by 
taking into account all the dimension-4 operators in the OPE.
The lines correspond
to the cases of $T_{PS}^{\rm Born}(0)$ $=0$, with (case (i))
and without (case (ii)) 
an explicit Born term ($T_{PS}^{\rm Born}(0)$ $\neq 0$) 
in the ``ph''side respectively. 
The remaining three lines show the results obtained
by extracting the contribution of
$m_c\langle\bar{q}q\rangle_N$ term in the OPE
from the above each line respectively.}
\label{fig:1}
\end{figure}

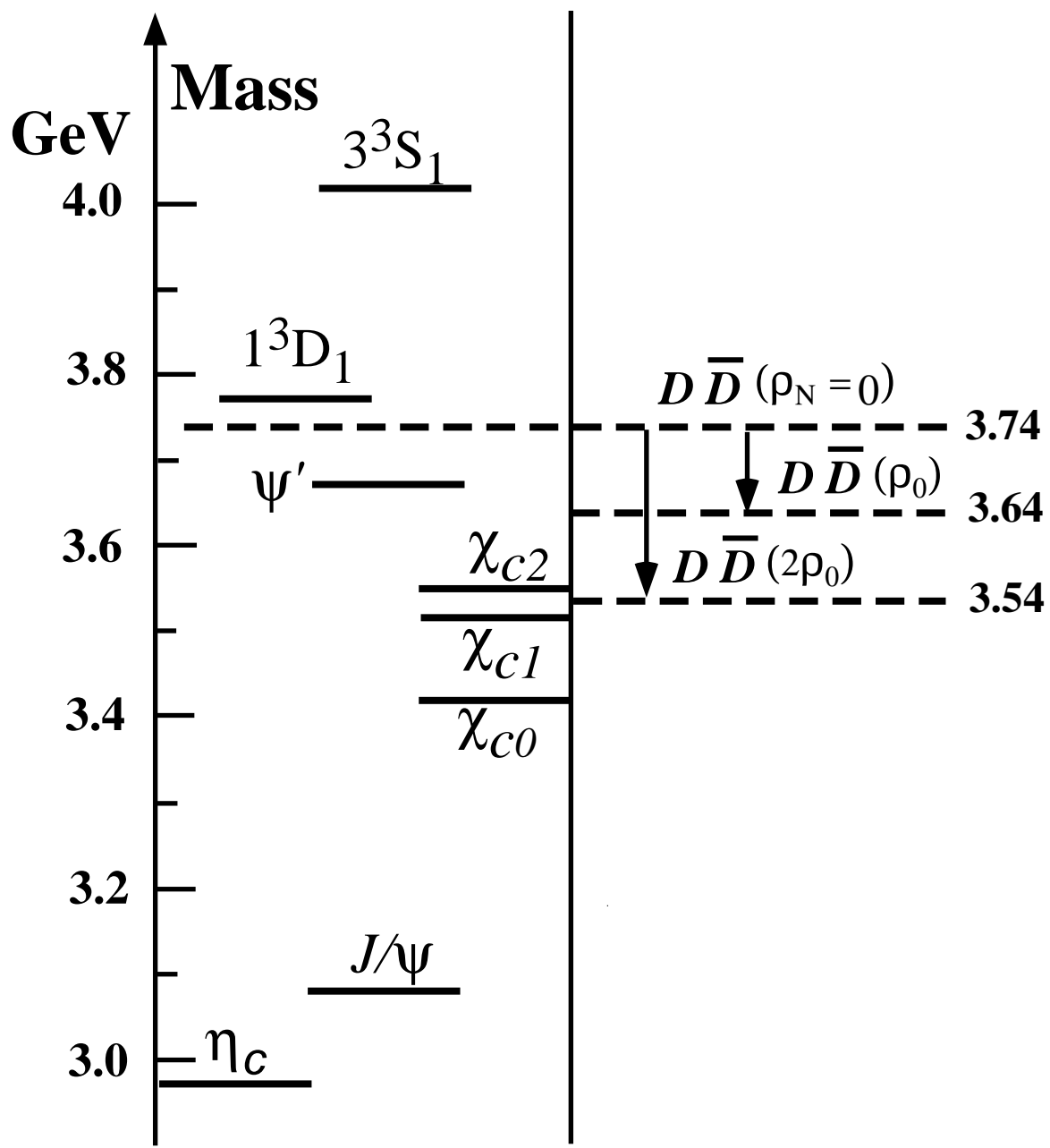
\begin{figure}
%\begin{center}
%\epsfile{file=charmspectrum.eps, scale=1.5}
%\end{center}
\caption{Comparison between charmonim spectrum
and $D\bar{D}$ threshold level:
If the mass shifts of all the charmonium states
are very small ($0.1\sim 0.2\%$ of the
bare mass) in the nuclear medium 
($\rho_N=\rho_0\sim 2\rho_0$, $\rho_0=0.17$ fm$^{-3}$), 
we predict that the threshold lying just below $1D$ state 
in free space falls down below the $\psi'$ state at normal matter density
and the $\chi_{c2}$ state twice at the density.}
\label{fig:2}
\end{figure}

\begin{center}
\begin{table}[h]
\caption{The result of an analysis for the solid,
dashed and doted lines in Fig.1:
The second raw shows the plateau region, which
we determine 
as the deviation from a minimum value of $\delta m_D$
in each line becomes less than $10\%$ of the minimum value.
The $\delta m_D$ in the plateau region is given in the third raw.
}
\begin{tabular}{cccc}
   Lines & 
   Solid line &
   Doted line (case (i)) &
   Dashed line (case (ii)) \\ \hline\hline
Plateau region [GeV$^2$] &
$3.5$ $\leq$ $M^2$ $\leq$ $7.0$ &
$2.6$ $\leq$ $M^2$ $\leq$ $7.0$ &
$3.4$ $\leq$ $M^2$ $\leq$ $8.0$ \\ \hline
$\delta m_D$ [GeV] &
$-63\pm 3$ &
$-53\pm 3$ &
$-44\pm 3$ \\
\end{tabular}
\label{table:1}
\end{table}
\end{center}


\begin{references}
\bibitem{QM97} Quark Matter '97,\ Proceeding of the 13th International
Conference on Ultra-Relativistic Nucleus-Nucleus Collisions, Tsukuba,
Japan, 1-5 Dec. 1997,\ {\it Nucl.\ Phys.} {\bf A638} (1998) 1c--610c.

\bibitem{BR} G.E. Brown and M. Rho,\ 
{\it Phys.\ Rev.\ Lett.} {\bf C66} (1991) 2720;\\
H. Siomi and T. Hatsuda, {\it Phys.\ Lett.} {\bf B334} (1994) 281;\\
M. Herrmann, B.L. Friman and W. N$\ddot{\mbox{o}}$renberg,\
{\it Nucl.\ Phys.} {\bf A560} (1993) 411;\\
M. Asakawa, C.M. Ko, P. L\'evai and X.J. Qiu,\ 
{\it Phys.\ Rev.} {\bf C46} (1992) R1159.

\bibitem{HL} T. Hatsuda and S.H. Lee,\
{\it Phys.\ Rev.} {\bf C46} (1992) R34.

\bibitem{YK} Y.\,Koike,\ {\it Phys.\ Rev.} {\bf C51} (1995) 1488.

\bibitem{JL} X. Jin and D.B. Leinweber,\ 
{\it Phys.\ Rev.} {\bf C52} (1995) 3344.

\bibitem{KH} Y. Koike and A. Hayashigaki,\ 
{\it Prog.\ Theo.\ Phys.} {\bf 98} (1997) 631.

\bibitem{FKW} F. Klingl, N. Kaiser and W. Weise,\
{\it Nucl.\ Phys.} {\bf A624} (1997) 527.

\bibitem{FHL}
R.J. Furnstahl, T. Hatsuda and S.H. Lee,
{\it Phys.\ Rev.} {\bf D42} (1990) 1744.

\bibitem{HKL} T. Hatsuda, Y. Koike and S.H. Lee,\
{\it Nucl.\ Phys.} {\bf B394} (1993) 221.

\bibitem{AH1}
A. Hayashigaki, {\it Prog.\ Theo.\ Phys.} {\bf 101} (1999) 923.

\bibitem{KKLMW}
F. Klingl, S. Kim, S.H. Lee, P. Morath
and W. Weise, {\it Phys.\ Rev.\ Lett.} {\bf 82} (1999) 3396.

\bibitem{CFG} T.D. Cohen, R.J. Furnstahl and D.K. Griegel,\ 
{\it Phys.\ Rev.} {\bf C45} (1992) 1881.

\bibitem{JCFG} X. Jin, T.D. Cohen, R.J. Furnstahl and D.K. Griegel,\ 
{\it Phys.\ Rev.} {\bf C47} (1993) 2882.

\bibitem{TLTSL} K. Tsushima, D.H. Lu, A.W. Thomas and R.H. Landau, 
{\it Phys.\ Rev.} {\bf C59} (1999) 2824;
A. Sibirtsev, K. Tsushima and A.W. Thomas,
{\it Eur.\ Phys.\ J.} {\bf A6} (1999) 351.

\bibitem{Alves}
G.A. Alves et al., {\it Phys.\ Rev.\ Lett.} {\bf 70} (1993) 722;\\
M.J. Leitch et al., {\it Phys.\ Rev.\ Lett.} {\bf 72} (1994) 2542.

\bibitem{DL} E.G. Drukarev and E.M. Levin, 
{\it Prog.\ Part.\ Nucl.\ Phys.} {\bf A556} (1991) 467

\bibitem{NN} F.S. Navarra and M. Nielsen,\ 	
{\it Phys.\ Lett.} {\bf B443} (1998) 285.

\bibitem{AH2} A. Hayashigaki,
The detailed explanation for the calculation
of Wilson coefficients is in preparation.

\bibitem{AE}
T.M. Aliev and V.L. Eletskii,\  
{\it Sov.\ J.\ Nucl.\ Phys.} {\bf 38}(6) (1983) 936.

\bibitem{RRY} L.J. Reinders, H.R. Rubinstein and Y. Yazaki,\ 
{\it Phys.\ Lett.} {\bf 97B} (1980) 257.

\bibitem{EKMRS}
A.X. El-Khadra, A.S. Kronfeld, P.B. Mackenzie,
S.M. Ryan and J.N. Simone,\ {\it Phys.\ Rev.} {\bf D58} (1998) 014506.

\bibitem{PDG}
Particle Data Group,\ {\it Eur.\ Phys.\ J.} {\bf C3} (1998) 353.

\bibitem{KMN} Y. Kondo, O. Morimatsu and Y. Nishino,
{\it Phys.\ Rev.} {\bf C53} (1996) 1927.

\bibitem{YNMK} H. Yabu, S. Nakamura, F. Myhrer and K. Kubodera,\ 
{\it Phys.\ Lett.} {\bf B315} (1993) 17.

\bibitem{RRY2} L.J. Reinders, H.R. Rubinstein and Y. Yazaki,\ 
{\it Phys.\ Rep.} {\bf 127} (1985) 1.

\bibitem{NA50} L. Rammello et al. (NA50 Collaboration), 
{\it Nucl.\ Phys.} {\bf A638} (1998) 261c; \\
M.C. Abreu et al. (NA50 Collaboration), 
{\it Phys.\ Lett.} {\bf 450B} (1999) 456; \\
C. Cical\'o, QUARK MATTER 99, Torino Italy, May 10-15, 1999,\\  
http://www.qm99.to.infn.it/cicalo/cicalo.html.

\bibitem{MS} T. Matsui and H. Satz, 
{\it Phys.\ Lett.} {\bf 178B} (1986) 416.

\bibitem{Anto} L. Antoniazzi et al. (E705 Collaboration), 
{\it Phys.\ Rev.\ Lett.} {\bf 70} (1993) 383;\\
Y.Lemoigne et al.,\ {\it Phys.\ Lett.} {\bf 113B} (1982) 509.

\bibitem{KNS} D. Kharzeev, M. Nardi and H. Satz,\ {\it hep-ph}/9707308;\\ 
D. Kharzeev,\ {\it Nucl.\ Phys.} {\bf A638} (1998) 279c;\\
R. Vogt, {\it Phys.\ Rep.} {\bf 310} (1999) 197. 

\bibitem{AH3} A. Hayashigaki,
A comparison beween our results
and the experimental data of NA50 Collaboration is under investigation.

\bibitem{VJ} R. Vogt and A. Jackson,\ 
{\it Phys.\ Lett.} {\bf 206B} (1988) 333.

\bibitem{STST} A. Sibirtsev, K. Tsushima, K. Saito and A.W. Thomas, 
{\it nucl-th}/9904015.

\bibitem{Kar}
D. Kharzeev and H. Satz,\ {\it Phys.\ Lett.} {\bf B334} (1994) 155;
{\it Phys.\ Lett.} {\bf B356} (1995) 365. 

\end{references}
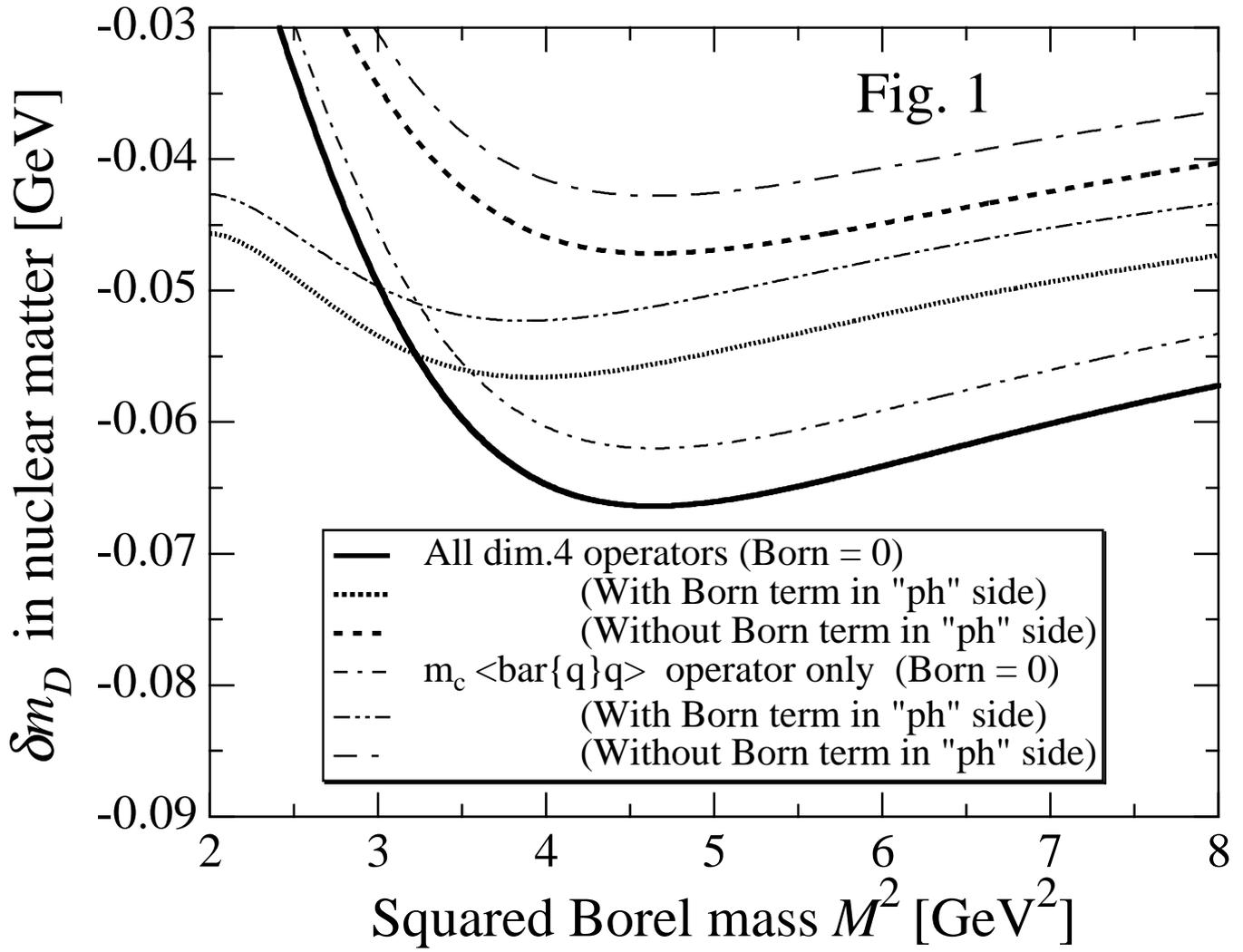
\end{document}